\documentclass[sigchi-a,nonacm]{acmart}
\usepackage[utf8]{inputenc}

\AtBeginDocument{%
  \providecommand\BibTeX{{%
    \normalfont B\kern-0.5em{\scshape i\kern-0.25em b}\kern-0.8em\TeX}}}

\setcopyright{acmcopyright}
\copyrightyear{2018}
\acmYear{2018}
\acmDOI{10.1145/1122445.1122456}

\copyrightyear{2020}
\acmYear{2020}
\acmConference[CHI '21]{CHI '21: ACM CHI Conference on Human Factors in Computing Systems}{May 8--13, 2021}{Yokohama, Japan}
\acmBooktitle{CHI '21: ACM CHI Conference on Human Factors in Computing Systems,
  May 8--1321, 2021, Yokohama, Japan}
\acmPrice{15.00}
\acmISBN{978-1-4503-XXXX-X/18/06}

\definecolor{dkgreen}{RGB}{0, 180, 0}

\begin{document}

\title{Conversational User Interfaces As Assistive interlocutors For Young Children's Bilingual Language Acquisition}


\author{Neelma Bhatti}
\affiliation{%
  \institution{Department of Computer Science}
  \city{Virginia Tech, Blacksburg}
  \state{USA}
}
\email{neelma@vt.edu}

\author{Timothy L. Stelter}
\affiliation{%
  \institution{Department of Computer Science}
  \city{Virginia Tech, Blacksburg}
  \state{USA}
}
\email{tstelter@vt.edu}

\author{D. Scott McCrickard}
\affiliation{%
  \institution{Department of Computer Science}
  \city{Virginia Tech, Blacksburg}
  \state{USA}
}
\email{mccricks@vt.edu}

\renewcommand{\shortauthors}{Neelma Bhatti et al.}

\begin{abstract}
Children in a large number of international and cross-cultural families in and outside of the US learn and speak more than one language. However, parents often struggle to acquaint their young children with their local language if the child spends majority of time at home and with their spoken language if they go to daycare or school. By reviewing relevant literature about the role of screen media content in young children's language learning, and interviewing a subset of parents raising multilingual children, we explore the potential of designing conversational user interfaces which can double as an assistive language aid. We present a preliminary list of objectives to guide the the design of conversational user interfaces dialogue for young children's bilingual language acquisition.
\end{abstract}



\maketitle

\marginpar{\begin{math} ^1 \end{math} Wikipedia defines Conversational User Interfaces (CUIs) as "a user interface for computers that emulates a conversation with a real human [...] (which) provides opportunity for the user to communicate with the computer in their natural language rather than in a syntax specific commands" \cite{CUIwiki}. Different terms appear in the literature for various types of CUIs including voice assistants \cite{xu2019young,beneteau2020parenting}, intelligent personal assistants \cite{wu2020see,clark2019makes}, smart speakers \cite{beneteau2020parenting}, voice user interfaces \cite{murad2019don}, voice based conversational agents \cite{lovato2019hey} and conversational agents \cite{bhatti2020interactive,clark2019makes}. In this paper, we use the terms CUI and CA interchangeably while referring to voice based conversational agents which are commonly used in domestic settings such as Google Home, Amazon Alexa, Echo Dot, Siri etc.}

\section{Introduction}
A large number of international parents both in and outside of the US raise children who speak more than one language. These parent(s) usually have minimal spoken communication with or in the presence of their children, and they usually do not have co-located extended family members, who can converse with the children using their native language. Additionally, such children feel challenged to communicate with their age mates in a language different from the one spoken at their home. Research shows that parents often use screen media content to acquaint their children with their parent's native language and to also help them become proficient in the language of communication in the country that they reside in\cite{beneteau2020parenting}. However, screen media content with animated characters (popular examples:\textit{Dora the Explorer, Mickey Mouse Clubhouse}) as well as human characters (popular examples: \textit{Blippi, Mother Goose Club, Blue's Clues}) provide a form of interactivity and incidental language learning for young children primarily in English language only.

Children's communication with conversational user interfaces (CUI)\begin{math} ^1 \end{math}  is a relatively nascent domain of research \cite{beneteau2020parenting,lovato2019hey,druga2017hey}. Conventional conversational partners for children (both in the form of voice interfaces and screen media characters) are unresponsive, do not give the children the feeling of being understood, and are incapable of maintaining an uninterrupted and seamless conversational flow \cite{bhatti2020interactive}. Children also struggle communicating with conversational agents (CA) due to grammar and language complexity \cite{beneteau2020parenting}. Although parents seek opportunities for using CUIs to expand their children's communication skills and learning another language \cite{beneteau2020parenting}, the functionality of switching languages is not available in state-of-the art CUIs. We explore this understudied domain by investigating the capability of CUIs as an interactive language aid for young children (2-5 year old) that can help children learn a secondary language in the absence of an interlocutor. We leverage the qualities of screen media in aiding children with language learning, and try to translate those qualities into the design of CUI for children.


\section{Approach}
Having a conversation partner in a true sense has a different meaning to children than adults, as attributes of conversation which are valued by adults such as trustworthiness and establishing common ground are not relevant to young children \cite{clark2019makes}. Instead, children value familiarity \cite{howard2013building, troseth2018let} and interactivity \cite{bhatti2020interactive} offered by screen media characters which are not a part of present day CUIs. Therefore, in the pursuit of proposing design goals for CUI which can act as an assistive language aid for children, we adopted a three-pronged approach while focusing on the aspects of dialogue that need to be \textit{different} from the conversational flow with adults. We first reviewed relevant literature on children's interaction with voice agents. Next, we conducted interviews with twelve mothers in the US who belonged to a subset of immigrant and non-US parents raising young bilingual children. Finally, a group of three researchers met to discuss and encapsulate the design goals of a CUI for young children in an ideation session \cite{bhatti2020interactive} based on the literature, requirements of the user population, and the experience of first author as a mother as well as a keen observer of her bilingual preschooler's screen media watching sessions.  



\section{Design Goals}
In the subsequent sections, we present a list of objectives (not arranged by importance) for the design of CUIs as assistive interlocutors for young children's bilingual language acquisition.


\subsection{Language Chunks}
Children of four of the mothers we interviewed experienced speech delay due to the cognitive load of understanding and communicating in more than one language. One of the four mothers repeatedly expressed how screen media assisted her child's speech more than sessions with his speech therapist or play-dates with his age mates, while another mother shared her child's speech therapist's strategy which seemed to be effective for her son:

\begin{quote}
    \textit{"I was having a lot of difficulty because we were using very long sentences to communicate with our child, like we used with each other, so as a result he (child) did not answer us. She (speech therapist) showed me how to use smaller words and sentences with the child so that he understands them and responds"}
\end{quote}

Short and succinct utterances by intelligent personal assistants have been found to be helpful for non-native language speakers \cite{wu2020see}. The importance of short sentences and common, easily graspable words has also been reported in literature, as lengthy and complex utterances by CA can make children vary of the responses \cite{lovato2019hey}. Hence, design of CA as assistive interlocutor could benefit from responding to a child differently than it does to an adult by \textit{communicating in simple, decomposed language}.

\subsection{Familiar Language Scaffolding}
Children of different ages and developmental delays need varying levels of language scaffolds \cite{dieterich2006impact}. While socially contingent video viewing sessions can effectively promote preschoolers science learning \cite{xu2020wonder}, research suggests the insufficiency of  communicative social cues to support toddler's word learning via video presentation in the absence of a co-present communication scaffold \cite{troseth2018let} such as verbal scaffolding by mothers of 3 to 4 year old children, which can result in richer verbal interactions and better language skills \cite{skibbe2004parental}. Thus, we believe that training the CA with attributes of language such as familiar voice, talking style, and commonly used words from the child's native language can help it mimic the child's family members including parents or older siblings. We take into account that parents of such children who have not lived in their native place for an extended period of time might not have good grip on the language themselves. Hence we also consider training the CUI in a remote location with familiar language scaffolding before being deployed in the home. Despite the critiques of mimicry objective \cite{aylett2019siri}, we argue for familiar language scaffolding due to evidence in literature \cite{howard2013building, troseth2018let} about the positive effects of children's para-social relationship with media characters on their learning due to the familiarity offered by them.

\marginpar{
\section{Design goals at a glance}
\begin{enumerate}
    \item \textbf{Language chunks:} communicating in simple, decomposed language
    \item \textbf{Familiar language scaffolding:} training the conversational agent with familiar voice, talking style, and commonly used words from their native language 
       \item \textbf{Contextual appropriateness:} contextually appropriate, where context can be a child’s age, mood, or developmental milestones
     \item \textbf{Social communication cues:} secondary language speaking interlocutor in a multi-person dialogue
    \item \textbf{Code switching and accent variations:} attuning the conversation to children's code switching and accent variations

\end{enumerate}
}



\subsection{Contextual Appropriateness}
Young children have a limited attention span and are more prone to distractions than adults; for instance, disruptions in the flow of conversation are likely to shift their attention elsewhere. The state of the art screen media shows do a good job of grabbing children's attention through interactive chatter and questioning. As one of the mothers mentioned:

\begin{quote}
    \textit{"Blippi is kind of like Dora, except that instead of waiting after asking question, he just goes and plays by himself. So I think they (children) get attracted to him because it’s not that he’s always asking questions, he just asks a question or says hi and waits for the response, and then he does something or starts playing. I think asking questions is kind of their way of grasping the attention of the viewer".}
\end{quote} 
 
As screen media offers interactivity and learning for children with human and animated characters attracting children's attention by asking them questions and performing activities that can keep them engaged, we envision this asynchronous interactivity in CUI's to be beneficial for children's language learning. Thus for engaging young children in meaningful dialogue, it is important for a conversational agent to be designed with a goal of being \textit{contextually appropriate}, where context can be a child's age, mood, or developmental milestones.

\subsection{Social Communication Cues}
Several interviews participants mentioned their children learning English as a secondary language by watching static content (rhymes, cartoons) and shows with characters which urge children to respond to them (\textit{Blippi, Dora, Mickey Mouse Clubhouse}). Children as young as 2 year old can learn new words by being a passive part of a social interaction, where they can hear two individuals conversing with each other \cite{o2011third}. Additionally, on screen social cues have  been reported to be associated with greater child engagement \cite{troseth2018let}. In consideration of attributes of screen media and normal social interactions which can successfully engage and acquaint children with new words, we envision a scenario where CA can stage a show with different voice characters having conservation among themselves while also involving children to enable language learning without assistance from family members. This phenomenon was also acknowledged by the first author, who co-watched \textit{Dora the Explorer} with her preschooler, where the titular character would occasionally rub off Spanish words in a primarily English show by giving the impression that certain characters in the show don't understand English. The portrayal of one of characters as a \textit{secondary language speaking interlocutor in a multi-person dialogue} urges the children to understand and communicate in a different language. Apart from helping the child learn the language, it can also enable them to understand multi-person social cues.



\subsection{Code Switching And Accent Variations}
Young children tend to use unclear or goofy tone when communicating with CAs as they find it to be a fun activity \cite{beneteau2020parenting}, making it difficult for the CA to answer children due to incorrect syntax, mispronunciation \cite{lovato2019hey}, or mixed utterances \cite{cantone2007code}. Multilingual children may have an accent that is typically different from accents supported by the CA. They also tend to mix words from different languages (code-switching) as they are unable to have a handle on switching between languages due to their young age \cite{yow2018code}. While conventional screen media content provides age-appropriate and partially interactive communication sessions \cite{xu2020wonder}, contemporary voice agents are not attuned to child friendly conversations \cite{beneteau2020parenting}. Thus, we propose \textit{attuning the conversation to children's code switching and accent variations} as a design goal in order to maintain a seamless conversation flow.


\section{Future Directions}
Despite the popularity of CUIs in domestic settings \cite{beneteau2020parenting, lovato2019hey}, their potential to be used as assistive interlocutors to learn a secondary language has not been fully realized. With our initial explorations in the domain of CUIs as assistive interlocutors for young children, we present a preliminary, yet prioritized list of high-level objectives which can guide the design of CUIs to aid bilingual children's secondary language learning. The application of such a CUI is beyond language learning by bilingual children, and can branch out to learning a language as a part of curricular requirement for older children. Due to the restrictions of scope and page limit, our proposed design goals are relevant to CUIs which are used exclusively by children, hence we do not account for situations which can have multiple people from the same household interacting with them \cite{huang2020amazon}. Future work in this domain should investigate the impact of culture on the interaction of bilingual children with a CA \cite{lovato2019hey}. Further research can also investigate the prospects of using CUI in conjunction with screen media to promote a greater level of engagement of children with the virtual interlocutor.



\bibliographystyle{ACM-Reference-Format}
\bibliography{sample-base}

\end{document}